\documentclass[12pt]{article}
\usepackage{geometry}                
\usepackage{graphicx}
\usepackage{amssymb}
\usepackage{epstopdf}
\usepackage{color}

\usepackage{amsmath}
\usepackage{graphics,graphicx,epsfig}
\usepackage{amssymb}
\usepackage{sidecap}
\sidecaptionvpos{figure}{c}
\usepackage{appendix}
\usepackage{cite}

\titlepage
\title{Birkhoff's Theorem in Lovelock Gravity for General Base Manifolds}

\usepackage{geometry}                
\usepackage{graphicx}
\usepackage{amssymb}
\usepackage{epstopdf}
\usepackage{color}

\usepackage{amsmath}
\usepackage{graphics,graphicx,epsfig}
\usepackage{amssymb}
\usepackage{sidecap}
\sidecaptionvpos{figure}{c}
\usepackage{appendix}

\begin{document}

\begin{titlepage}
\vfill
\vskip 3.0cm
\begin{center}
{\Large\bf Birkhoff's Theorem in Lovelock Gravity for General Base Manifolds}

\vskip 10.mm
{\bf  Sourya Ray} 
\vskip 0.1 in Instituto de Ciencias F\'{\i}sicas y Matem\'{a}ticas, Universidad Austral de
Chile, Valdivia, Chile\\
\vskip 0.1 in Email: \texttt{ray@uach.cl}\\
\vspace{0.5 in}
\end{center}

\begin{center}
{\bf Abstract}
\end{center}

\begin{quote}
We extend the Birkhoff's theorem in Lovelock gravity for arbitrary base manifolds using an elementary method. In particular, it is shown that any solution of the form of a warped product of a two-dimensional transverse space and an arbitrary base manifold must be static. Moreover, the field equations restrict the base manifold such that all the non-trivial intrinsic Lovelock tensors of the base manifold are constants, which can be chosen arbitrarily, and the metric in the transverse space is determined by a single function of a spacelike coordinate which satisfies an algebraic equation involving the constants characterizing the base manifold along with the coupling constants.
   \vfill
\vskip 2.mm
\end{quote}
\hfill
\end{titlepage}
\section{Introduction}
Birkhoff's theorem in general relativity states that the only spherically symmetric vacuum solution of Einstein's equations in four spacetime dimensions is the Schwarzschild solution. This theorem is also valid in any dimensions $D>4$, where the corresponding solution is the higher dimensional analogue known as the Schwarzschild-Tangherlini solution. In order to prove this theorem, one starts with a general metric ansatz for a spherically symmetric spacetime
\begin{align}
ds^2=-f(r,t)dt^2+\dfrac{dr^2}{g(r,t)}+r^2d\Sigma_{D-2}^2,
\label{metansatz}
\end{align}
where $d{\Sigma}_{D-2}^2=\hat{g}_{\mu\nu}dx^{\mu}dx^{\nu}$ is the line element of a $(D-2)$-sphere, which hereafter we shall call the base manifold. The Einstein's field equations then (after a redefinition of the timelike coordinate $t$) uniquely determine the metric functions $f(r,t)$ and $g(r,t)$ in terms of a single function of the spacelike coordinate $r$:
\begin{align*}
f(r,t)=g(r,t)=1-\dfrac{M}{r^{D-3}}.
\end{align*}
Now, if the base manifold is left arbitrary in the metric ansatz, then the field equations only imply that it must be an (Euclidean) Einstein manifold and hence one can obtain another solution by replacing the round $(D-2)$-sphere by any other Einstein manifold with the same Ricci curvature. However, in four spacetime dimensions, the round $2$-sphere is the unique (up to discrete quotients) Einstein manifold with positive curvature. In higher dimensions there are other possibilities. Hence, Birkhoff's theorem in higher dimensions is valid for a wider class of metrics which in particular include spacetimes with the B\"{o}hm metrics existing on base manifolds which are topologically spheres or products thereof\footnote{Note however that the Schwarzschild-Tangherlini metric is the unique {\it asymptotically flat} spacetime in the whole family.}\cite{Bohm:1998,Gibbons:2002bh,Gibbons:2002av}.

Birkhoff's theorem also holds in Lovelock theory of gravity \cite{Zegers:2005vx,Deser:2005gr} which is a natural higher curvature generalization of Einstein's general relativity and retains the property that the field equations are of second order in the metric \cite{Lovelock:1971yv}. The Lovelock action is given by the Lagrangian density
\begin{align}
\mathcal{L}=\sqrt{g}\sum_{k=0}^{\bar{k}}c_k\mathcal{L}^{(k )}, \quad \quad \quad \mathcal{L}^{(k)}=\dfrac{1}{2^k}\delta_{a_1b_1\cdots a_kb_k}^{c_1d_1\cdots c_kd_k}R_{c_1d_1}^{\ \ \ \ a_1b_1}\cdots R_{c_kd_k}^{\ \ \ \ a_kb_k}
\label{Lagr}
\end{align}
and the equations of motion are given by
\begin{align}
0=\mathcal{G}^a_c=\sqrt{g}\sum_{k=0}^{\bar{k}}c_k\mathcal{G}^{(k)a}_{\ \ \ c}, \quad \quad \quad \mathcal{G}^{(k)a}_{\ \ \ c}=-\dfrac{1}{2^{k+1}}\delta^{aa_1b_1\cdots a_kb_k}_{cc_1d_1\cdots c_kd_k}R^{\ \ \ \ c_1d_1}_{a_1b_1}\cdots R^{\ \ \ \ c_kd_k}_{a_kb_k}
\label{fieldeqns}
\end{align}
where $\delta^{a_1\cdots a_n}_{c_1\cdots c_n}=n!\delta^{a_1}_{[c_1}\cdots\delta^{a_n}_{c_n]}$ is the totally anti-symmetric generalized Kronecker delta and $c_k$ are the coupling constants corresponding to each term. The first and the second term in the summation in (\ref{Lagr}) corresponds to the cosmological constant and the Einstein-Hilbert term respectively. In any spacetime dimensions $D$, there can be at most a finite number of non-trivial terms in the sum. This is because beyond a certain order (equal to the integral part of $(D-1)/2$) the terms in the action are either identically zero or are topological invariants whose variation with respect to the metric gives a total derivative and consequently does not contribute to the equations of motion.

In this article, we shall extend the Birkhoff's theorem in Lovelock gravity for general base manifolds. Such spacetimes were earlier studied either in Gauss-Bonnet gravity (i.e., $\bar{k}=2$) and/or for particular non-maximally-symmetric base manifolds \cite{Barcelo:2002wz,Dotti:2005rc,Dotti:2008pp,Dotti:2010bw,Bogdanos:2010zz,Maeda:2010bu,
Bogdanos:2009pc,Oliva:2012ff,Pons:2014oya,Farhangkhah:2014zka,Dadhich:2015nua}. Our analysis applies to a general Lovelock gravity of arbitrary order with generic coupling constants, in arbitrary dimensions and for arbitrary base manifolds. To this end, we shall express the field equations in a compact form which will allow us to employ an elementary method to analyze them.

\section{Staticity and Reduced Field Equations}

We consider $D$-dimensional spacetimes given by the metrics of the form (\ref{metansatz}) but with arbitrary base manifolds.

In the following we shall use the latin letters at the beginning of the alphabet e.g $a, b, c,...$ to denote general spacetime indices and those somewhat at the middle of the alphabet like $i,j,k,...$ to denote spacetime indices on the two dimensional transverse part of the metric spanned by the coordinates $(t,r)$ while we shall use greek alphabets $\alpha, \beta, \gamma,...$ to denote spacetime indices on the $(D-2)$-dimensional base manifold. Also note that all quantities defined intrinsically on the base manifold will appear with a hat, for example $\hat{R}_{\alpha \beta}^{\ \ \mu\nu}$ will denote the intrinsic Riemann curvature components evaluated on the base manifold using the metric $\hat{g}_{\mu\nu}$. 

After carrying out some algebra, one finds that, {\it for generic values of the coupling constants $c_k$}, the $\mathcal{G}^r_t=\mathcal{G}^t_r=0$ components of the field equations imply that the Levi-Civita component $\Gamma^r_{rt}$ must vanish and consequently the metric function $g(r,t)$ must be independent of the timelike coordinate $t$ i.e., $g(r,t)=g(r)$. Furthermore, the component $\mathcal{G}^r_r-\mathcal{G}^t_t=0$ of the field equations will then imply that the metric function $f(r,t)$ must be of the form $\kappa(t)g(r)$ where $\kappa(t)$ is an arbitrary function of $t$ which can be set to unity by redefining the coordinate $t$. This implies that the general solution of the form (\ref{metansatz}) of the field equations (\ref{fieldeqns}) must be {\it static} and reduces to
\begin{align}
ds^2=-g(r)dt^2+\dfrac{dr^2}{g(r)}+r^2d\Sigma_{D-2}^2
\label{redmetansatz}
\end{align}
This reduced metric ansatz simplifies the remaining field equations considerably which can then be expressed as
\begin{align}
0=\mathcal{G}^i_j=-\dfrac{\sqrt{g}\ \delta^i_j}{2(D-1)!r^{D-2}}\sum\limits_{n=0}^{\bar{k}}\left\{(D-2n-2)!\hat{\mathcal{L}}^{(n)}\right\}\left\{\dfrac{d\ }{dr}\left(r^{D-2n-1}A_n(U)\right)\right\}
\label{redfeqns1}
\end{align}
and
\begin{align}
0=\mathcal{G}^{\alpha}_{\beta}=\dfrac{\sqrt{g}}{(D-1)!r^{D-3}}\sum\limits_{n=0}^{\bar{k}}\left\{(D-2n-3)!\hat{\mathcal{G}}^{(n)\alpha}_{\ \ \ \beta}\right\}\left\{\dfrac{d^2\ }{dr^2}\left(r^{D-2n-1}A_n(U)\right)\right\}
\label{redfeqns2}
\end{align}
where $\hat{\mathcal{L}}^{(n)}$ and $\hat{\mathcal{G}}^{(n)\alpha}_{\ \ \ \beta}$ are the Euler invariants and the Lovelock tensors of order $n$ of the base manifold. The functions $A_n(U)$ are polynomials of order $\bar{k}-n$ given by
\begin{align*}
A_n(U)=\sum\limits_{k=n}^{\bar{k}}a_k \binom{k}{n}U^{k-n}=\sum\limits_{p=0}^{\bar{k}-n}a_{n+p}\binom{n+p}{n}U^p
\end{align*}
where $a_k$ are rescaled coupling constants given by
\begin{align*}
a_k=\dfrac{(D-1)!}{(D-2k-1)!}c_k
\end{align*}
and $U=-g(r)/r^2$. Note that the polynomials $A_n(U)$ satisfy the recurrence relations
\begin{align*}
A_n'(U)=(n+1)A_{n+1}(U).
\end{align*}
From the expressions (\ref{redfeqns1}) and (\ref{redfeqns2}), we see that the reduced field equations are of the form
\begin{align*}
\sum\limits_{k}f_k(x)g_k(y)=0
\end{align*}
 which is a functional equation of {\it Pexider type} and can be solved in an elementary way \cite{Aczel:2006fe,Suto:1914}. The basic idea is that for a non-trivial solution to exist, the sets of functions $\{f_k(x)\}$ and $\{g_k(y)\}$ must be linearly dependent sets. 
To see this, one just needs to successively differentiate the equation with respect to each of the variables $k$ times and thereby obtain two systems of equations, each linear in either set of functions. One can then see that the ranks of the matrices corresponding to each system of equations (which are actually the Wronskian matrices of the two sets of functions) must add up to the number of functions in each set i.e., the range of $k$. In particular if one set of functions is a linearly independent set then all the functions in the other set must be zero.
 
\section{Constraints on the Base Manifold and Generalized Wheeler Polynomial} 
 We now use the above method to solve the reduced field equations (\ref{redfeqns1}) and (\ref{redfeqns2}). As mentioned previously, the highest order term in the action contributes to the field equations only for $D>2\bar{k}$. The analysis then can be divided into three separate cases. For the sake of simplicity we first consider
 
 \textbf{Case I:} \underline{$D>2\bar{k}+2$:} Let us consider the equation (\ref{redfeqns1}) first. Since the polynomials $A_n(U)$ depend on the coupling constants, which are arbitrarily specified, the Wronskian of the functions 
 \begin{align*}
 \psi_n=\dfrac{d\ }{dr}\left(r^{D-2n-1}A_n(U)\right) \qquad \qquad \forall \ n=0,1,\ldots,\bar{k}
 \end{align*}
 must be of maximal\footnote{Obviously, if the rank of the Wronskian is $\bar{k}+1$ then the set of functions $\lbrace \psi_n \rbrace$ is linearly independent which implies that $\hat{\mathcal{L}}^{(n)}=0$ for all values of $n$, which contradicts the fact $\hat{\mathcal{L}}^{(0)}=1$.} rank $\bar{k}$, so that there is a {\it single} equation of the form
\begin{align}
\sum\limits_{n=0}^{\bar{k}}\alpha_n \psi_n=0
\label{lindep}
\end{align} 
 which can be trivially integrated to
\begin{align}
\sum\limits_{n=0}^{\bar{k}}\alpha_nr^{D-2n-1}A_n(U)=\text{constant}
\label{genwheelerpoly}
\end{align}
which in turn determines the unknown function $g(r)$. This is the generalization of the polynomial equation, first given by Wheeler \cite{Wheeler:1985qd}, whose solution determines the metric function for spherically symmetric solutions of Lovelock gravity. However, the constants $\alpha_n$'s generally depend on the geometry of the base manifold as we shall show now.

Once the rank of the Wronskian of the set of functions $\{\psi_n\}$ is set to $\bar{k}$, the rank of the Wronskian of the set of functions $\{(D-2n-2)!\hat{\mathcal{L}}^{(n)}\}$ must be equal to $1$ (since there are a total of $\bar{k}+1$ number of functions in each set). In other words, all the Euler invariants of the base manifold must be multiples of each other. Moreover, since $\hat{\mathcal{L}}^{(0)}=1$, they all must also be constants. In fact, they are given by
\begin{align}
\hat{\mathcal{L}}^{(n)}=\dfrac{q\alpha_n}{(D-2n-2)!}
\label{Eulerdenst}
\end{align}
where $q$ is an arbitrary multiplicative factor which can be fixed using $\hat{\mathcal{L}}^{(0)}=1$, which gives us $q=(D-2)!$, setting $\alpha_0=1$.

Next we solve the field equations (\ref{redfeqns2}) with components on the base manifold. As before, for arbitrary coupling constants, the rank of the Wronskian of the set of functions $\left\lbrace\dfrac{d\psi_n}{dr}\right\rbrace$ must be $\bar{k}$, in which case there is a single relation among the functions given by
\begin{align*}
\sum\limits_{n=0}^{\bar{k}}\alpha_n \dfrac{d\psi_n}{dr}=0.
\end{align*} 
The coefficients here are set to $\alpha_n$ in order to be consistent with (\ref{lindep}). This in turn implies that the Lovelock tensors on the base manifold must be given by
\begin{align}
\hat{\mathcal{G}}^{(n)\alpha}_{\ \ \ \beta}=-\dfrac{1}{2(D-2)}\dfrac{q\alpha_n}{(D-2n-3)!}\delta^{\alpha}_{\beta}=-\dfrac{(D-3)!\alpha_n}{2(D-2n-3)!}\delta^{\alpha}_{\beta}.
\label{Lovlocktens}
\end{align} 
where the overall factor is so chosen such that
\begin{align*}
\hat{\mathcal{G}}^{(n)\alpha}_{\ \ \ \alpha}=-\dfrac{(D-2n-2)}{2}\hat{\mathcal{L}}^{(n)}.
\end{align*}
In summary, for generic coupling constants, the solutions of the form (\ref{metansatz}) are given by a single metric function $g(r)$ which solves the algebraic equation (\ref{genwheelerpoly}), such that the Lovelock tensors on the base manifold are given by (\ref{Lovlocktens}). Finally, we consider  

\textbf{Case II:} \underline{$D=2\bar{k}+1$:} In this case the summations in equations (\ref{redfeqns1}) and (\ref{redfeqns2}) extend up to $n=\bar{k}-1$, since both the Euler invariant $\hat{\mathcal{L}}^{(\bar{k})}$ and the Lovelock tensor $\hat{\mathcal{G}}^{(\bar{k})\alpha}_{\ \ \ \beta}$ identically vanishes on the base manifold and $\psi_{\bar{k}}=0$. This in turn implies that the summation in (\ref{genwheelerpoly}) which determines the metric function $g(r)$ also ranges from $0$ to $\bar{k}-1$ and the base manifold is restricted by constant Lovelock tensors up to order $\bar{k}-1$ by (\ref{Lovlocktens}).

\textbf{Case III:} \underline{$D=2\bar{k}+2$:} In this case the summations in equations (\ref{redfeqns1}) and (\ref{redfeqns2}) extend up to $n=\bar{k}$ and $n=\bar{k}-1$ respectively, while the summation in (\ref{genwheelerpoly}) ranges from $0$ to $\bar{k}$ and the base manifold is restricted by constant Lovelock tensors up to order $\bar{k}-1$ by (\ref{Lovlocktens}) and the Euler invariant $\hat{\mathcal{L}}^{(\bar{k})}=(2\bar{k})!\alpha_{\bar{k}}$ which is related to the Euler class for compact base manifolds without any boundary.

The algebraic equation (\ref{genwheelerpoly}) can be solved explicitly for $g(r)$ only up to $\bar{k}=4$ for generic values of the coupling constants. We now simplify the algebraic equation for two special choices of the coupling constants and then show that the results are in agreement with the known ones for constant curvature base manifolds \cite{Crisostomo:2000bb}.

{\bf Solution of Maximum Multiplicity:} There can be at most $\bar{k}$ distinct solutions to the equation (\ref{genwheelerpoly}) depending on the coupling constants $a_n$ and the constants characterizing the base manifold $\alpha_n$. In this case we assume that all the constants are such that the algebraic equation has a unique solution of maximum multiplicity. We first rewrite the equation as follows
\begin{align*}
\sum\limits_{n=0}^{\bar{k}}\dfrac{\alpha_n}{r^{2n}}A_n(U)=\dfrac{\text{constant}}{r^{D-1}}
\end{align*}
We note that the left hand side is an inhomogeneous polynomial in $U$ and $1/r^2$ of degree $\bar{k}$. Furthermore, observing the highest order term in $U$, we infer that the equation has a unique solution if and only if the left hand side can be expressed as
\begin{align*}
a_{\bar{k}}\left(U+\beta+\dfrac{\gamma}{r^2}\right)^{\bar{k}}
\end{align*}
Expanding and then equating both expressions we obtain the following relation among the constants
\begin{align*}
\alpha_na_m=a_{\bar{k}}\binom{\bar{k}}{m}\beta^{\bar{k}-m}\gamma^n \qquad \text{for} \:\: 0\leqslant n\leqslant m\leqslant \bar{k}
\end{align*}
Putting $m=\bar{k}$ we conclude that all the coupling constants $a_n$ and those characterizing the base manifold $\alpha_n$ must be related to the parameters $\beta$ and $\gamma$ through the relations
\begin{align*}
\alpha_n=\gamma^n \qquad \text{and} \qquad a_k=a_{\bar{k}}\binom{\bar{k}}{k}\beta^{\bar{k}-k},
\end{align*}

Hence the metric function in this case is given by
\begin{align*}
g(r)=-r^2U=\beta r^2+\gamma-\dfrac{\text{const.}}{r^{(D-2\bar{k}-1)/ \bar{k}}}
\end{align*}
However, as mentioned earlier, for this choice of the coupling constants the theory does not admit Birkhoff's theorem in that there exists more general solutions of the form (\ref{metansatz}).

{\bf Pure Lovelock Gravity:} Let us now analyze the case of pure Lovelock gravities, where there is a single term in the Lagrangian of order $\bar{k}$ with coupling $a_{\bar{k}}\neq 0$. The polynomial $A_n(U)$ then reduces to a monomial of order $\bar{k}-n$ and the equation (\ref{genwheelerpoly}) to
\begin{align*}
\sum\limits_{n=0}^{\bar{k}}\alpha_n\binom{\bar{k}}{n}(r^{2}U)^{\bar{k}-n}=\dfrac{\text{const.}}{r^{D-2\bar{k}-1}}
\end{align*} 
which generically leads to $\bar{k}$ distinct solutions for the metric function $g(r)=-r^2U$, each corresponding to the distinct roots of the polynomial equation. Moreover, if all the constants $\alpha_n$ for $n=1,\ldots,\bar{k}$ are related to a single parameter $\gamma$ by
\begin{align*}
\alpha_n=\gamma^n
\end{align*}
then the equation has a unique solution 
of the form
\begin{align*}
g(r)=-r^2U=\gamma-\dfrac{\text{const.}}{r^{(D-2\bar{k}-1)/ \bar{k}}}.
\end{align*}
\section{Conclusion}
In conclusion, we have extended the Birkhoff's theorem in Lovelock gravity for non-constant curvature base manifolds. Using a simple method we have shown that for generic coupling constants, the transverse part of the metric is given in terms of a single function of the spacelike coordinate $r$ which is determined by an algebraic equation. Furthermore, the field equations are shown to dictate that all the Lovelock tensors and the Euler invariants up to order $\bar{k}$ of the base manifold are constants, where $\bar{k}$ is the highest order of the curvature terms in the action. These constants (if not identically zero) can be chosen arbitrarily but once fixed they do appear in the metric function in the transverse space since they must satisfy the algebraic equation (\ref{genwheelerpoly}). It would be interesting to see how far do these constraints go in specifying the intrinsic curvature on the base manifold, particularly in dimensions where there are a maximal number of constraints. Moreover, generically the constants $\alpha_n$ characterizing the base manifold, along with the coupling constants $c_n$, determine the asymptotic structure of the corresponding solution. Hence, it is natural to wonder if these constants may have any physical interpretation. However, it is also important to note that for certain choices of the coupling constants, the field equations reduce to an under-determined set of equations and Birkhoff's theorem is no longer valid. These choices may even allow non-static solutions \cite{Dotti:2010bw,Oliva:2012ff}. It would also be nice to perform a complete classification of these ``degenerate" cases. Another straightforward extension of this work would be to include an electric charge in the analysis. Work along some of these lines are currently in progress. Finally, for Einstein base manifolds 
\begin{align*}
\hat{C}^{\ \ \ \ \rho\sigma}_{\alpha\beta}=\hat{R}^{\ \ \ \ \rho\sigma}_{\alpha\beta}-\alpha_1\delta^{\rho\sigma}_{\alpha\beta}
\end{align*}
and hence the constancy of all the Lovelock tensors and the Euler invariants can also be expressed in terms of the constancy of the tensors
\begin{align*}
\dfrac{1}{2^{k+1}}\delta^{\alpha \alpha_1\beta_1\cdots \alpha_k\beta_k}_{\rho\rho_1\sigma_1\cdots \rho_k\sigma_k}\hat{C}^{\ \ \ \ \rho_1\sigma_1}_{\alpha_1\beta_1}\cdots \hat{C}^{\ \ \ \ \rho_k\sigma_k}_{\alpha_k\beta_k}
&=\dfrac{(D-3)!\delta^{\alpha}_{\rho}}{2(D-2k-3)!}\sum\limits_{n=0}^k\binom{k}{n}(-\alpha_1)^n\alpha_{k-n}.
\end{align*}
where $\hat{C}_{\alpha\beta\rho\sigma}$ is the conformal (Weyl) tensor on the base manifold. One can check that our results are in agreement with the previously obtained results for the particular cases.

{\bf Acknowledgements:} Useful discussions with David Kastor, Julio Oliva and Jennie Traschen are gratefully acknowledged. This work is supported by FONDECYT grant 1150907.


\end{document}